\documentclass[fleqn,twoside]{article}
\usepackage{espcrc1}

\newcommand{\AmS}{{\protect\the\textfont2
  A\kern-.1667em\lower.5ex\hbox{M}\kern-.125emS}}

\title{Data allocation on disks
 with solution reconfiguration
 (problems, heuristics)
}

\author{Mark Sh. Levin
\thanks{
 Mark Sh. Levin:~
 Inst. for Inform. Transmission Problems,
 Russian Academy of Sciences;
  http://www.mslevin.iitp.ru;
 email: mslevin@acm.org
  }
  }

\begin{document}

\maketitle

\begin{abstract}
 The paper addresses
 problem of data allocation in two-layer computer storage
 while taking into account
 dynamic digraph(s) over computing tasks.
 The basic version of
  data file allocation on parallel hard magnetic disks
 is considered as special bin packing model.
 Two problems of the allocation solution reconfiguration
 (restructuring)
 are suggested:
 (i) one-stage restructuring model,
 (ii) multistage restructuring models.
 Solving schemes are based on simplified heuristics.
%
 Numerical examples illustrate problems and solving schemes.

~~


\vspace{1pc}
\end{abstract}

\maketitle

\tableofcontents

\newcounter{cms}
\setlength{\unitlength}{1mm}

\section{Introduction}

%
 In management/planning of hierarchical, distributed
  computer systems,
  problems of tasks/data placement in storage have been studied
 many years as allocation of objects (tasks, jobs, balls, data
 files) into set of resources (e.g., servers, computers, machines,
 bins, urns)
 (e.g., \cite{aven72,batu12,golub09,shach09,shach12,wolf97}).
%
%
%
%
 Mathematical modeling of the problems is often based on stochastic models
 (e.g., Markov processes)
 (e.g., \cite{golub09})
 and
 combinatorial optimization models
 (e.g.,
  multiple knapsack problems,
  location/assignment models,
  bin packing problems)
 (e.g., \cite{shach09,shach12}.
%
%
 One of the data placement problem is targeted to
  file allocation on
 a hard magnetic disk with moving disk driver heads
 (e.g., \cite{bor77,gur78,gur74,levm79}).
%
%
 Usually, the study of this kind of problems
 (as control of two-level storage)
 is based on the following approaches:
 (a) stochastic approach
  (e.g., \cite{bor77,gur78,gur74}),
 (b approximation solving schemes
   (e.g., \cite{golub09,shach09});
 (c) heuristic and metaheuristic solving schemes
  (e.g., \cite{bor77,gur78,gur74}).


%
 In this paper,
 problem of data file allocation in two-layer computer memory
 and
 parallel memories (disks) at the second layer
 while taking into account
 dynamic digraph(s) over computing tasks.
 The author version of file allocation on hard magnetic disks
 is examined as a special version of bin packing problem.
 This allocation problem is considered as the basic one.
 In addition, two optimization problems as
 reconfiguration of allocation solution(s) are examined:
 (i) one-stage restructuring
  \cite{lev11restr,lev15restr,shieb12},
 (ii) multistage restructuring
 \cite{lev11restr,lev15restr}.
%
%
 Some basic simple heuristic ideas are described and
 corresponding simplified solving schemes are used.
 A numerical example
 illustrates the file allocation problems,
 reconfiguration of allocation solutions,
 and simple solving schemes (heuristics).

\section{General problems types}

 Our generalized description of the considered problem
 is the following (Fig. 1):

~

   {\bf Problem :} \( < M_{proc}  -\alpha | M_{o}-\beta | M_{e}-\gamma  >\)

~

  where
   \( M_{proc}\) corresponds to processors
  (\( \alpha \) is the number of parallel processors);
 \(M_{o}\) corresponds to operation memory
  (\(\beta\) is the number of parallel operation memories);
 \(M_{e}\) corresponds to external memory
  (\(\gamma \) is the number of parallel external memories, e.g.,
  disks).

 Generally, the following six
  basic computer hierarchy cases can be examined:


 (a) {\bf Problem 1:} \(1\) processor,  \(1\) memory,  \(1\) disk
 ~\(< M_{proc} -1 | M_{o}-1 | M_{e}-1 >\) (Fig. 2a)
%
 (e.g., \cite{lev81,lev15});

 (b) {\bf Problem 2:} \(1\) processor,  \(1\) memory,  \(\gamma\) disks
 ~\(< M_{proc} -1 | M_{o}-1 | M_{e}- \gamma >\) (Fig. 2b)
 (e.g, \cite{levm79});

 (c) {\bf Problem 3:} \(1\) processor,
 \(\beta\) memory,  \(\gamma\) disks
 ~\(< M_{proc}  -1 | M_{o}-\beta | M_{e} -\gamma >\) (Fig. 2c, \(\beta = \gamma\));

 (d) {\bf Problem 4:} \(\alpha\) processors,  \(1\) memory,   \(1\) disk
 ~\(< M_{proc} -\alpha | M_{o}-1 | M_{e}-1>\) (Fig. 3a);
%

 (e) {\bf Problem 5:} \(\alpha\) processors, \(\beta\) memories,  \(1\) disk
 ~\( < M_{proc} -\alpha | M_{o}-\beta | M_{e}-1 >\)
 (Fig. 3b, \(\alpha = \beta \));

 (f) {\bf Problem 6:} \(\alpha\) processors,  \(\beta\) memories,  \(\gamma\) disks
 ~\( < M_{proc}-\alpha | M_{o}-\beta | M_{e}-\gamma >\)

 (Fig. 3c, \(\alpha = \beta = \gamma \) ).

\begin{center}
\begin{picture}(100,50)
\put(23.5,00){\makebox(0,0)[bl] {Fig. 1. General  frameworks}}

\put(04,38){\makebox(0,0)[bl]{Complicated computing}}
\put(04,34){\makebox(0,0)[bl]{tasks system}}
\put(04,30){\makebox(0,0)[bl]{(e.g., dynamic digraph}}
\put(04,26){\makebox(0,0)[bl]{over computing tasks)}}

\put(23,34){\oval(46,19)} \put(23,34){\oval(45,18)}


\put(05,19){\vector(0,1){05}} \put(11,19){\vector(0,1){05}}
\put(17,19){\vector(0,1){05}} \put(23,19){\vector(0,1){05}}
\put(29,19){\vector(0,1){05}} \put(35,19){\vector(0,1){05}}
\put(41,19){\vector(0,1){05}}

\put(05,23){\vector(0,-1){05}} \put(11,23){\vector(0,-1){05}}
\put(17,23){\vector(0,-1){05}} \put(23,23){\vector(0,-1){05}}
\put(29,23){\vector(0,-1){05}} \put(35,23){\vector(0,-1){05}}
\put(41,23){\vector(0,-1){05}}


\put(14.5,11){\makebox(0,0)[bl]{Data (files)}}

\put(00,09){\line(1,0){46}} \put(00,17){\line(1,0){46}}
\put(00,09){\line(0,1){08}} \put(46,09){\line(0,1){08}}

\put(0.5,09){\line(0,1){08}} \put(45.5,09){\line(0,1){08}}

\put(51,42){\makebox(0,0)[bl]{Mapping}}

\put(55,38){\makebox(0,0)[bl]{\(\Longrightarrow \)}}
\put(55,34){\makebox(0,0)[bl]{\(\Longrightarrow \)}}

\put(55,30){\makebox(0,0)[bl]{\(\Longrightarrow \)}}
\put(55,26){\makebox(0,0)[bl]{\(\Longrightarrow \)}}
\put(55,22){\makebox(0,0)[bl]{\(\Longrightarrow \)}}
\put(55,18){\makebox(0,0)[bl]{\(\Longrightarrow \)}}
\put(55,14){\makebox(0,0)[bl]{\(\Longrightarrow \)}}
\put(55,10){\makebox(0,0)[bl]{\(\Longrightarrow \)}}

\put(78,43.5){\makebox(0,0)[bl]{Processor}}
\put(80,40){\makebox(0,0)[bl]{system}}

\put(70,39){\line(1,0){30}} \put(70,47){\line(1,0){30}}
\put(70,39){\line(0,1){08}} \put(100,39){\line(0,1){08}}


\put(85,38){\vector(0,-1){4.4}} \put(85,34){\vector(0,1){5}}

\put(70.5,28.6){\makebox(0,0)[bl]{Operation memory }}

\put(85,30){\oval(30,07)}


\put(85,26){\vector(0,-1){4.4}} \put(85,22){\vector(0,1){4.4}}

\put(71,14){\makebox(0,0)[bl]{External memory }}
\put(71,10){\makebox(0,0)[bl]{(e.g., disks)}}

\put(70,07){\line(0,1){13}} \put(100,07){\line(0,1){13}}

\put(85,20){\oval(30,03)} \put(85,07){\oval(30,03)}

\end{picture}
\end{center}

\begin{center}

\begin{picture}(30,61)

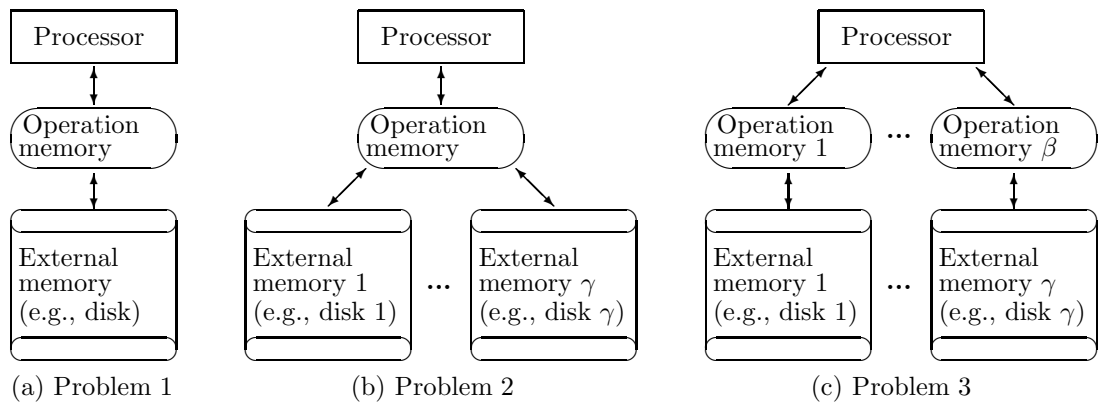
\put(37,00){\makebox(0,0)[bl] {Fig. 2. One-processor problems
 1, 2, and 3}}

\put(00,05){\makebox(0,0)[bl] {(a) Problem 1}}

\put(03,52.5){\makebox(0,0)[bl]{Processor}}

\put(00,50){\line(1,0){22}} \put(00,57){\line(1,0){22}}
\put(00,50){\line(0,1){07}} \put(22,50){\line(0,1){07}}


\put(11,49.5){\vector(0,-1){5}} \put(11,44.5){\vector(0,1){5}}

\put(01.5,40){\makebox(0,0)[bl]{Operation}}
\put(01,37){\makebox(0,0)[bl]{memory }}

\put(11,40){\oval(22,08)}


\put(11,35){\vector(0,-1){4.4}} \put(11,31){\vector(0,1){4.4}}

\put(01,23){\makebox(0,0)[bl]{External}}
\put(01,19){\makebox(0,0)[bl]{memory }}
\put(01,15){\makebox(0,0)[bl]{(e.g., disk)}}

\put(00,12){\line(0,1){17}} \put(22,12){\line(0,1){17}}

\put(11,29){\oval(22,03)} \put(11,12){\oval(22,03)}



\end{picture}
%
\begin{picture}(60,59)


\put(14,05){\makebox(0,0)[bl] {(b) Problem 2}}

\put(18,52.5){\makebox(0,0)[bl]{Processor}}

\put(15,50){\line(1,0){22}} \put(15,57){\line(1,0){22}}
\put(15,50){\line(0,1){07}} \put(37,50){\line(0,1){07}}


\put(26,49.5){\vector(0,-1){5}} \put(26,44.5){\vector(0,1){5}}

\put(16.5,40){\makebox(0,0)[bl]{Operation}}
\put(16,37){\makebox(0,0)[bl]{memory}}

\put(26,40){\oval(22,08)}


\put(16,36){\vector(-1,-1){5}} \put(11,31){\vector(1,1){5}}

\put(01,23){\makebox(0,0)[bl]{External}}
\put(01,19){\makebox(0,0)[bl]{memory \(1\)}}
\put(01,15){\makebox(0,0)[bl]{(e.g., disk \(1\))}}

\put(00,12){\line(0,1){17}} \put(22,12){\line(0,1){17}}

\put(11,29){\oval(22,03)} \put(11,12){\oval(22,03)}


\put(24,20){\makebox(0,0)[bl]{{\bf ...}}}



\put(36,36){\vector(1,-1){5}} \put(41,31){\vector(-1,1){5}}

\put(31,23){\makebox(0,0)[bl]{External}}
\put(31,19){\makebox(0,0)[bl]{memory \(\gamma\)}}
\put(31,15){\makebox(0,0)[bl]{(e.g., disk \(\gamma\))}}

\put(30,12){\line(0,1){17}} \put(52,12){\line(0,1){17}}

\put(41,29){\oval(22,03)} \put(41,12){\oval(22,03)}

\end{picture}
%
\begin{picture}(52,59)


\put(14,05){\makebox(0,0)[bl] {(c) Problem 3}}

\put(18,52.5){\makebox(0,0)[bl]{Processor}}

\put(15,50){\line(1,0){22}} \put(15,57){\line(1,0){22}}
\put(15,50){\line(0,1){07}} \put(37,50){\line(0,1){07}}


\put(16,49.5){\vector(-1,-1){5}} \put(11,44.5){\vector(1,1){5}}

\put(01.5,40){\makebox(0,0)[bl]{Operation}}
\put(01,37){\makebox(0,0)[bl]{memory \(1\)}}

\put(11,40){\oval(22,08)}


\put(11,35){\vector(0,-1){4.4}} \put(11,31){\vector(0,1){4.4}}

\put(01,23){\makebox(0,0)[bl]{External}}
\put(01,19){\makebox(0,0)[bl]{memory \(1\)}}
\put(01,15){\makebox(0,0)[bl]{(e.g., disk \(1\))}}

\put(00,12){\line(0,1){17}} \put(22,12){\line(0,1){17}}

\put(11,29){\oval(22,03)} \put(11,12){\oval(22,03)}


\put(24,40){\makebox(0,0)[bl]{{\bf ...}}}
\put(24,20){\makebox(0,0)[bl]{{\bf ...}}}


\put(36,49.5){\vector(1,-1){5}} \put(41,44.5){\vector(-1,1){5}}

\put(31.5,40){\makebox(0,0)[bl]{Operation}}
\put(31,37){\makebox(0,0)[bl]{memory \(\beta\)}}

\put(41,40){\oval(22,08)}


\put(41,35){\vector(0,-1){4.4}} \put(41,31){\vector(0,1){4.4}}

\put(31,23){\makebox(0,0)[bl]{External}}
\put(31,19){\makebox(0,0)[bl]{memory \(\gamma\)}}
\put(31,15){\makebox(0,0)[bl]{(e.g., disk \(\gamma\))}}

\put(30,12){\line(0,1){17}} \put(52,12){\line(0,1){17}}

\put(41,29){\oval(22,03)} \put(41,12){\oval(22,03)}

\end{picture}
\end{center}



 Fig. 4 illustrates one-stage allocation of data files on hard magnetic disks
 as bin packing problem.
 The following designations are used:
 (a)  digraph \(D = (T,R )\),
  where \(T\) is the set of computing tasks,
  \(R\) is precedence relation as a set of arcs over the
  computing tasks above;
  (b) data files and processing graph over them
 \(G = <Q,E_{1},E_{2} >\),
 where \(Q\) is the set of data files under processing,
  \(E_{1}\) is precedence binary relation over the files
   (i.e., a set of arcs),
 \(E_{2}\) is symmetric binary relation of common processing
 of data files
 (i.e., concurrently,
   a set of edges).


\begin{center}
\begin{picture}(50,59)

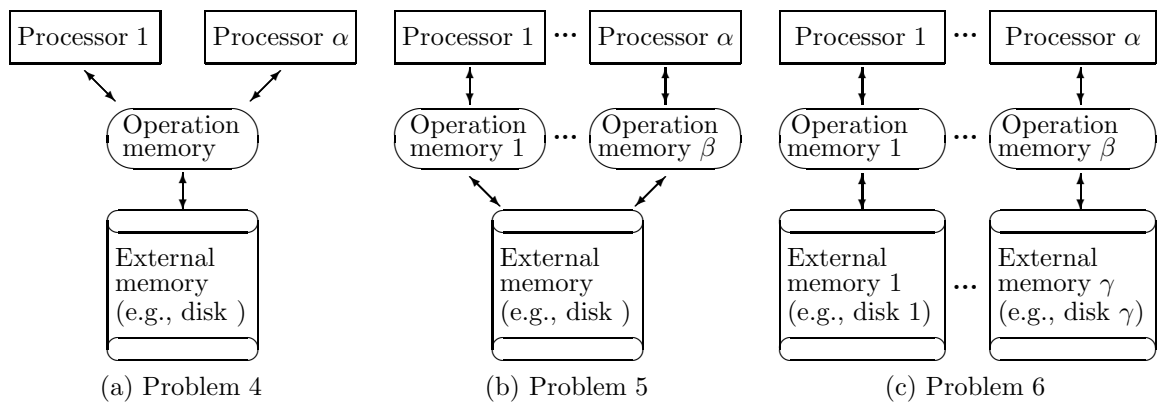
\put(41,00){\makebox(0,0)[bl] {Fig. 3. Multi-processor
  problems 4, 5, and 6}}

\put(12,05){\makebox(0,0)[bl] {(a) Problem 4}}

\put(01,52.5){\makebox(0,0)[bl]{Processor \(1\)}}

\put(00,50){\line(1,0){20}} \put(00,57){\line(1,0){20}}
\put(00,50){\line(0,1){07}} \put(20,50){\line(0,1){07}}

\put(10,49){\vector(1,-1){4}} \put(14,45){\vector(-1,1){4}}




\put(27,52.5){\makebox(0,0)[bl]{Processor \(\alpha\)}}

\put(26,50){\line(1,0){20}} \put(26,57){\line(1,0){20}}
\put(26,50){\line(0,1){07}} \put(46,50){\line(0,1){07}}


\put(36,49){\vector(-1,-1){4}} \put(32,45){\vector(1,1){4}}

\put(015,40){\makebox(0,0)[bl]{Operation}}
\put(015,37){\makebox(0,0)[bl]{memory }}

\put(23,40){\oval(20,08)}


\put(23,34.5){\vector(0,-1){4}} \put(23,31.5){\vector(0,1){4}}

\put(014,23){\makebox(0,0)[bl]{External}}
\put(014,19){\makebox(0,0)[bl]{memory }}
\put(014,15){\makebox(0,0)[bl]{(e.g., disk )}}

\put(13,12){\line(0,1){17}} \put(33,12){\line(0,1){17}}

\put(23,29){\oval(20,03)} \put(23,12){\oval(20,03)}








\end{picture}
\begin{picture}(50,59)


\put(12,05){\makebox(0,0)[bl] {(b) Problem 5}}

\put(01,52.5){\makebox(0,0)[bl]{Processor \(1\)}}

\put(00,50){\line(1,0){20}} \put(00,57){\line(1,0){20}}
\put(00,50){\line(0,1){07}} \put(20,50){\line(0,1){07}}

\put(10,49){\vector(0,-1){4.4}} \put(10,45){\vector(0,1){4.4}}


\put(21,53.5){\makebox(0,0)[bl]{{\bf ...}}}

\put(27,52.5){\makebox(0,0)[bl]{Processor \(\alpha\)}}

\put(26,50){\line(1,0){20}} \put(26,57){\line(1,0){20}}
\put(26,50){\line(0,1){07}} \put(46,50){\line(0,1){07}}


\put(36,49){\vector(0,-1){4.4}} \put(36,45){\vector(0,1){4.4}}

\put(02,40){\makebox(0,0)[bl]{Operation}}
\put(02,37){\makebox(0,0)[bl]{memory \(1\)}}

\put(10,40){\oval(20,08)}


\put(10,35){\vector(1,-1){4}} \put(14,31){\vector(-1,1){4}}

\put(014,23){\makebox(0,0)[bl]{External}}
\put(014,19){\makebox(0,0)[bl]{memory }}
\put(014,15){\makebox(0,0)[bl]{(e.g., disk )}}

\put(13,12){\line(0,1){17}} \put(33,12){\line(0,1){17}}

\put(23,29){\oval(20,03)} \put(23,12){\oval(20,03)}


\put(21,40){\makebox(0,0)[bl]{{\bf ...}}}

\put(27.5,40){\makebox(0,0)[bl]{Operation}}
\put(27,37){\makebox(0,0)[bl]{memory \(\beta\)}}

\put(36,40){\oval(20,08)}


\put(36,35){\vector(-1,-1){4}} \put(32,31){\vector(1,1){4}}

\end{picture}
\begin{picture}(50,59)


\put(14,05){\makebox(0,0)[bl] {(c) Problem 6}}

\put(02,52.5){\makebox(0,0)[bl]{Processor \(1\)}}

\put(00,50){\line(1,0){22}} \put(00,57){\line(1,0){22}}
\put(00,50){\line(0,1){07}} \put(22,50){\line(0,1){07}}

\put(11,49){\vector(0,-1){4.4}} \put(11,45){\vector(0,1){4.4}}


\put(23,53.5){\makebox(0,0)[bl]{{\bf ...}}}

\put(30,52.5){\makebox(0,0)[bl]{Processor \(\alpha\)}}

\put(28,50){\line(1,0){22}} \put(28,57){\line(1,0){22}}
\put(28,50){\line(0,1){07}} \put(50,50){\line(0,1){07}}


\put(40,49){\vector(0,-1){4.4}} \put(40,45){\vector(0,1){4.4}}


\put(01.5,40){\makebox(0,0)[bl]{Operation}}
\put(01,37){\makebox(0,0)[bl]{memory \(1\)}}

\put(11,40){\oval(22,08)}


\put(11,35){\vector(0,-1){4.4}} \put(11,31){\vector(0,1){4.4}}

\put(01,23){\makebox(0,0)[bl]{External}}
\put(01,19){\makebox(0,0)[bl]{memory \(1\)}}
\put(01,15){\makebox(0,0)[bl]{(e.g., disk \(1\))}}

\put(00,12){\line(0,1){17}} \put(22,12){\line(0,1){17}}

\put(11,29){\oval(22,03)} \put(11,12){\oval(22,03)}


\put(23,40){\makebox(0,0)[bl]{{\bf ...}}}
\put(23,20){\makebox(0,0)[bl]{{\bf ...}}}



\put(29.5,40){\makebox(0,0)[bl]{Operation}}
\put(29,37){\makebox(0,0)[bl]{memory \(\beta\)}}

\put(39,40){\oval(22,08)}


\put(40,35){\vector(0,-1){4.4}} \put(40,31){\vector(0,1){4.4}}

\put(29,23){\makebox(0,0)[bl]{External}}
\put(29,19){\makebox(0,0)[bl]{memory \(\gamma\)}}
\put(29,15){\makebox(0,0)[bl]{(e.g., disk \(\gamma\))}}

\put(28,12){\line(0,1){17}} \put(50,12){\line(0,1){17}}

\put(39,29){\oval(22,03)} \put(39,12){\oval(22,03)}

\end{picture}
\end{center}

\begin{center}
\begin{picture}(44,75)

\put(00,00){\makebox(0,0)[bl] {Fig. 4.
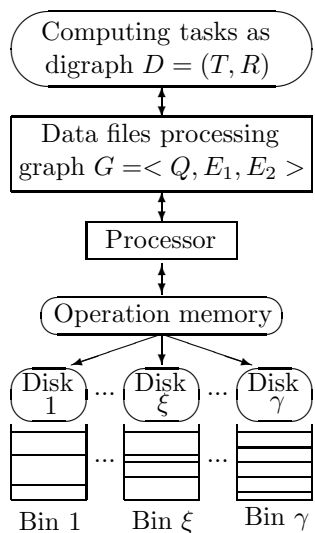
  File location on disks }}


\put(20,69){\oval(40,10)}

\put(4,69.5){\makebox(0,0)[bl]{Computing tasks as}}
\put(04.5,65){\makebox(0,0)[bl]{digraph
 \(D = (T,R )\)}}

\put(20,64){\vector(0,-1){4}} \put(20,60){\vector(0,1){4}}

\put(04,56){\makebox(0,0)[bl]{Data files processing}}
\put(01,51.5){\makebox(0,0)[bl]{graph \(G
 = <Q,E_{1},E_{2} >\)}}

\put(00,50){\line(1,0){40}} \put(00,60){\line(1,0){40}}
\put(00,50){\line(0,1){10}} \put(40,50){\line(0,1){10}}

\put(20,50){\vector(0,-1){4}} \put(20,46){\vector(0,1){4}}

\put(12.5,42.5){\makebox(0,0)[bl]{Processor }}

\put(10,41){\line(1,0){20}} \put(10,46){\line(1,0){20}}
\put(10,41){\line(0,1){05}} \put(30,41){\line(0,1){05}}

\put(20,40.4){\vector(0,-1){4}} \put(20,36.4){\vector(0,1){4}}

\put(05.6,32){\makebox(0,0)[bl]{Operation memory}}

\put(20,33.5){\oval(32,05)}

\put(20,31){\vector(-3,-1){12}} \put(20,31){\vector(0,-1){4}}
\put(20,31){\vector(3,-1){12}}

\put(05,23){\oval(10,07)}

\put(1.4,23.5){\makebox(0,0)[bl]{Disk }}
\put(4.4,20.5){\makebox(0,0)[bl]{\(1\)}}

\put(00,09){\line(1,0){10}}

\put(00,09){\line(0,1){10}} \put(10,09){\line(0,1){10}}

\put(00,11){\line(1,0){10}} \put(00,15){\line(1,0){10}}
\put(00,18){\line(1,0){10}}

\put(01,05){\makebox(0,0)[bl]{Bin \(1\) }}


\put(11,23){\makebox(0,0)[bl]{...}}

\put(11,14){\makebox(0,0)[bl]{...}}

\put(20,23){\oval(10,07)}

\put(16.4,23.5){\makebox(0,0)[bl]{Disk }}
\put(19.4,20.1){\makebox(0,0)[bl]{\(\xi\)}}

\put(15,09){\line(1,0){10}}

\put(15,09){\line(0,1){10}} \put(25,09){\line(0,1){10}}

\put(15,12){\line(1,0){10}} \put(15,14){\line(1,0){10}}
\put(15,15){\line(1,0){10}} \put(15,18){\line(1,0){10}}

\put(16,04.5){\makebox(0,0)[bl]{Bin \(\xi\) }}


\put(26,23){\makebox(0,0)[bl]{...}}

\put(26,14){\makebox(0,0)[bl]{...}}


\put(35,23){\oval(10,07)}

\put(31.4,23.5){\makebox(0,0)[bl]{Disk }}
\put(34.4,20.5){\makebox(0,0)[bl]{\(\gamma\)}}

\put(30,09){\line(1,0){10}}

\put(30,09){\line(0,1){10}} \put(40,09){\line(0,1){10}}

\put(30,10){\line(1,0){10}} \put(30,12){\line(1,0){10}}
\put(30,14){\line(1,0){10}} \put(30,16){\line(1,0){10}}
\put(30,18){\line(1,0){10}}

\put(31,05){\makebox(0,0)[bl]{Bin \(\gamma\) }}

\end{picture}
\end{center}

 Further, it is reasonable
 to examine  time sequence
 \(<t_{1},...,t_{j},..., t_{k}>\)
 and the corresponding sequence of computing tasks digraphs:~
 \( < D^{1} = (T^{1},R^{1}),..., D^{j} = (T^{j},R^{j}),...,
 D^{k} = (T^{k},R^{k})\)
 (Fig. 5).
 Evidently, the computing tasks digraph sequence requires
 allocation of data files on disks.

\begin{center}
\begin{picture}(120,45)

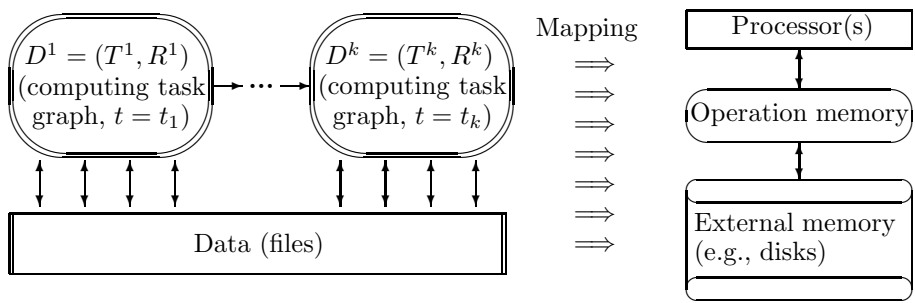
\put(14,00){\makebox(0,0)[bl] {Fig. 5. Processing a sequence of
 computing task digraphs}}

\put(02,36){\makebox(0,0)[bl]{\(D^{1}=(T^{1},R^{1})\)}}

\put(01,32){\makebox(0,0)[bl]{(computing task }}
\put(03,28){\makebox(0,0)[bl]{graph, \(t=t_{1}\))}}

\put(13.5,34){\oval(27,19)} \put(13.5,34){\oval(26,18)}


\put(04,19){\vector(0,1){05}} \put(10,19){\vector(0,1){05}}
\put(16,19){\vector(0,1){05}} \put(22,19){\vector(0,1){05}}

\put(04,23){\vector(0,-1){05}} \put(10,23){\vector(0,-1){05}}
\put(16,23){\vector(0,-1){05}} \put(22,23){\vector(0,-1){05}}


\put(27,34){\vector(1,0){04}}

\put(31.8,33.8){\makebox(0,0)[bl]{{\bf ... } }}

 \put(36,34){\vector(1,0){04}}

\put(42,36){\makebox(0,0)[bl]{\(D^{k}=(T^{k},R^{k})\)}}

\put(41,32){\makebox(0,0)[bl]{(computing task }}
\put(43,28){\makebox(0,0)[bl]{graph, \(t=t_{k}\))}}

\put(53.5,34){\oval(27,19)} \put(53.5,34){\oval(26,18)}


\put(44,19){\vector(0,1){05}} \put(50,19){\vector(0,1){05}}
\put(56,19){\vector(0,1){05}} \put(62,19){\vector(0,1){05}}

\put(44,23){\vector(0,-1){05}} \put(50,23){\vector(0,-1){05}}
\put(56,23){\vector(0,-1){05}} \put(62,23){\vector(0,-1){05}}


\put(24.5,11){\makebox(0,0)[bl]{Data (files)}}

\put(00,09){\line(1,0){66}} \put(00,17){\line(1,0){66}}
\put(00,09){\line(0,1){08}} \put(66,09){\line(0,1){08}}

\put(0.5,09){\line(0,1){08}} \put(65.5,09){\line(0,1){08}}

\put(70,40){\makebox(0,0)[bl]{Mapping}}

\put(75,36){\makebox(0,0)[bl]{\(\Longrightarrow \)}}
\put(75,32){\makebox(0,0)[bl]{\(\Longrightarrow \)}}

\put(75,28){\makebox(0,0)[bl]{\(\Longrightarrow \)}}
\put(75,24){\makebox(0,0)[bl]{\(\Longrightarrow \)}}
\put(75,20){\makebox(0,0)[bl]{\(\Longrightarrow \)}}
\put(75,16){\makebox(0,0)[bl]{\(\Longrightarrow \)}}
\put(75,12){\makebox(0,0)[bl]{\(\Longrightarrow \)}}

\put(96,40){\makebox(0,0)[bl]{Processor(s)}}

\put(90,39){\line(1,0){30}} \put(90,44){\line(1,0){30}}
\put(90,39){\line(0,1){05}} \put(120,39){\line(0,1){05}}


\put(105,38){\vector(0,-1){4.4}} \put(105,34){\vector(0,1){5}}

\put(90.5,28.6){\makebox(0,0)[bl]{Operation memory }}

\put(105,30){\oval(30,07)}


\put(105,26){\vector(0,-1){4.4}} \put(105,22){\vector(0,1){4.4}}

\put(91,14){\makebox(0,0)[bl]{External memory }}
\put(91,10){\makebox(0,0)[bl]{(e.g., disks)}}

\put(90,07){\line(0,1){13}} \put(120,07){\line(0,1){13}}

\put(105,20){\oval(30,03)} \put(105,07){\oval(30,03)}

\end{picture}
\end{center}


 The sequence of data files processing graphs is:
 \[ \overline{G} = < G^{1}=(Q^{1},E_{1}^{1} ,E_{2}^{1} ) \rightarrow ... \rightarrow
 G^{j}=(Q^{j},E_{1}^{j},E_{2}^{j}) \rightarrow ... \rightarrow
 G^{k}=(Q^{k},E_{1}^{k},E_{2}^{k}  )   >,\]
 where
 \(Q\) is the set of data files,
 \(E\) is the set of edges/arcs,
  \(G=(Q,E)\) is the general file processing graph,
 \(G^{j}=(A^{j},E_{1}^{j},E_{2}^{j}  )\) is the file procesisng  graph at time \(t_{j}\)
 \(A^{j} \subseteq A\),
 \(E_{1}^{j} \subseteq E_{1}\),
 \(E_{2}^{j} \subseteq E_{2}\).

 Note, the graph chains can be generalized to examine
 graph networks, e.g.,
 \(\overline{D} = (D,V)\),
 where \(D = \{ D^{j}, j=\overline{1,k}\} \) is the set of
 computing task digraphs,
 \(V\) is a set of arcs (i.e., precedence constraint
 over the set of computing task digraphs).
 Here many combinatorial optimization models
 can be used as auxiliary problems, for example
%
(e.g., \cite{gar79}):
 (i) multiple knapsack models,
%
 (ii) assignment/allocation models,
%
 (iii) bin packing models, and
%
 (iv) covering models.

\section{Allocation of files on hard magnetic disks}

%
%
%

\subsection{Problem statement}

 Our problem for data file allocation on hard magnetic disks
  has been suggested in
 \cite{levm79} as follows.
%
%
 Let
 \(Q = \{ 1,...,i,...,n \}\) be a set of data files,
 \(L = \{ 1,...,\xi,...,\gamma \}\)
 be a set of external memories (hard disks).
 Each disk \(\xi \in L\) has a number
 of free disk tracks
  \(W^{j}\) (i.e., disk size).
 The required memory size for each file
 (i.e., the required number of disk tracks)
 \(\forall i \in Q\) is: \(d_{i}\).
 Evidently, the global memory size constraint is:~
 \( \sum_{i=1}^{n} d_{i}  \leq  \sum_{\xi=1}^{\gamma} W^{\xi}  \).

 First, partitioning the files on disks is
 (without intersections):
 \(X = \{X_{1},...,X_{\xi},...,X_{\gamma}\}\)
 (\( | X_{\xi_{1}} \&  X_{\xi_{2}} | = 0 \), \( \forall \xi_{1}, \xi_{1} \in Q \) ),
 where
  set of files \(X_{\xi}\)
 (\(X_{\xi} \subseteq  Q\))
  is located on
 disk \(\xi\)
 and the size constraint for each disk \(\xi\) is:~
 \(\sum_{\kappa \in X_{\xi} } d_{\kappa} \leq W^{\xi}\),
 ~\( \forall \xi \in L\).
 In addition, at each disk \(\xi\) the correspponding files \(X_{\xi}\) are
 ordered  to get a linear ordering:
 \(\overline{X}_{\xi}\).
 Thus, the global solution is (file allocation):~
 \( \overline{X} =
 \{\overline{X}_{1},...,\overline{X}_{\xi},...,\overline{X}_{\gamma} \}\).

 Second,
 processing the files is defined by matrix movement probabilities
 (from one file \(i_{1}\) to another file  \(i_{2}\),
 this is defined by processing graph):~
%
 \( \Phi ( G ) = \|  \phi_{i_{1},i_{2}} \|^{n}_{i_{1},i_{2}=1} ,   ~~ i_{1},i_{2} \in Q\),
 where \( \phi_{i_{1},i_{2}}\) is a stationary probability of
 movement from file \(i_{1}\) to file \(i_{2}\)
 (in data file processing graph \(G\)).

%
 Let \(E_{3}\) be a symmetric binary relation of joint file processing
 (integration of \(E_{1}\) and \(E_{2}\)). For example,
 \(E_{3}\) can be defined by the rule:~
 \( ( (i_{1},i_{2}) \in E_{1} ) ~\bigcup~  ( (i_{1},i_{2}) \in E_{2} )
  \Rightarrow  (i_{1},i_{2}) \in E_{3}  ~~ \forall  i_{1},i_{2}
  \in Q\).

 Note location of files at different disks leads
 to concurrent processing the files without movement of disk drive head.
 Finally,
 the considered objective function
 for allocation of files \(\overline{X}\) is:
%
 \[ \min~ \Psi ( \overline{X} ) = \sum^{n}_{i_{1}\in X_{\xi_{1}},i_{2}\in X_{\xi_{2}}, \xi_{1} = \xi_{2}}
  ~  \phi_{i_{1},i_{2}} ~~ p_{i_{1},i_{2}} ( \overline{X} )  \]
 where
  \(p_{i_{1},i_{2}} (\overline{X})\) is a cost of disk drive head movement
 from file \(i_{1}\) to file \(i_{2}\)
 for solution \(\overline{X} \).

 Note, counting of
 \(p_{i_{1},i_{2}} (\overline{X})\)
 is a complicated problem and simplified methods are often applied.
%
%

 In Fig. 6,
  the problem of file re-allocation on disks is illustrated.

\begin{center}
\begin{picture}(100,85)

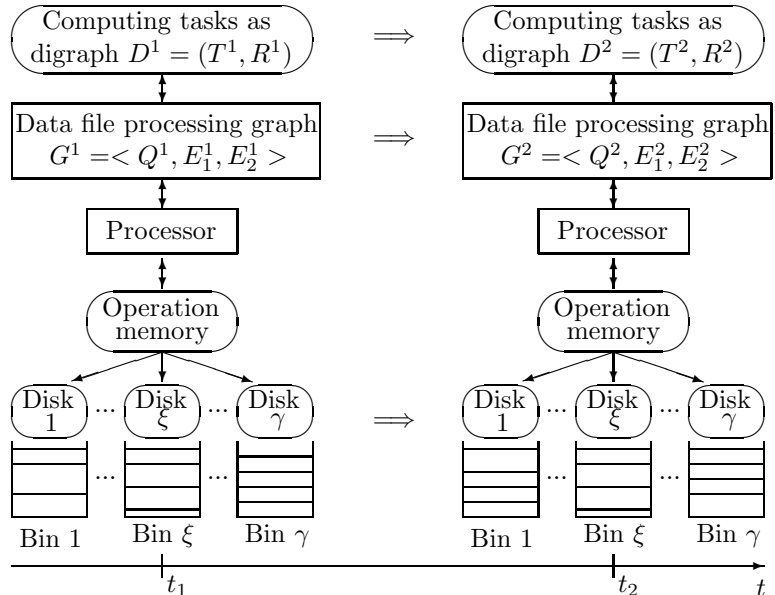
\put(06,00){\makebox(0,0)[bl] {Fig. 6. Illustration for
 re-allocation of data files on disks }}


\put(00,08.5){\vector(1,0){100}}

\put(20,7){\line(0,1){03}} \put(80,7){\line(0,1){03}}

\put(99,4.5){\makebox(0,0)[bl]{\(t\)}}

\put(20.6,04.5){\makebox(0,0)[bl]{\(t_{1}\)}}
\put(80.6,04.5){\makebox(0,0)[bl]{\(t_{2}\)}}


\put(20,78.5){\oval(40,09)}

\put(4,79.3){\makebox(0,0)[bl]{Computing tasks as}}
\put(02.5,74.7){\makebox(0,0)[bl]{digraph
 \(D^{1} = (T^{1},R^{1} )\)}}

\put(20,74){\vector(0,-1){4}} \put(20,70){\vector(0,1){4}}

\put(0.5,65.5){\makebox(0,0)[bl]{Data file processing graph}}
\put(05,61){\makebox(0,0)[bl]{\(G^{1}
 = <Q^{1},E_{1}^{1},E_{2}^{1} >\)}}

\put(00,60){\line(1,0){41}} \put(00,70){\line(1,0){41}}
\put(00,60){\line(0,1){10}} \put(41,60){\line(0,1){10}}

\put(20,60){\vector(0,-1){4}} \put(20,56){\vector(0,1){4}}

\put(12.5,52){\makebox(0,0)[bl]{Processor }}

\put(10,50){\line(1,0){20}} \put(10,56){\line(1,0){20}}
\put(10,50){\line(0,1){06}} \put(30,50){\line(0,1){06}}

\put(20,49.4){\vector(0,-1){4}} \put(20,45.4){\vector(0,1){4}}

\put(12,41){\makebox(0,0)[bl]{Operation}}
\put(13.8,38){\makebox(0,0)[bl]{memory }}

\put(20,41){\oval(20,08)}

\put(20,37){\vector(-3,-1){12}} \put(20,37){\vector(0,-1){4}}
\put(20,37){\vector(3,-1){12}}

\put(05,29){\oval(10,07)}

\put(1.4,29.5){\makebox(0,0)[bl]{Disk }}
\put(4.4,26.5){\makebox(0,0)[bl]{\(1\)}}

\put(00,15){\line(1,0){10}}

\put(00,15){\line(0,1){10}} \put(10,15){\line(0,1){10}}

\put(00,18){\line(1,0){10}} \put(00,22){\line(1,0){10}}
\put(00,24){\line(1,0){10}}

\put(01,11){\makebox(0,0)[bl]{Bin \(1\) }}


\put(11,29){\makebox(0,0)[bl]{...}}

\put(11,20){\makebox(0,0)[bl]{...}}

\put(20,29){\oval(10,07)}

\put(16.4,29.5){\makebox(0,0)[bl]{Disk }}
\put(19.4,26.5){\makebox(0,0)[bl]{\(\xi\)}}

\put(15,15){\line(1,0){10}}

\put(15,15){\line(0,1){10}} \put(25,15){\line(0,1){10}}

\put(15,16){\line(1,0){10}} \put(15,19){\line(1,0){10}}
\put(15,22){\line(1,0){10}} \put(15,24){\line(1,0){10}}

\put(16,11){\makebox(0,0)[bl]{Bin \(\xi\) }}


\put(26,29){\makebox(0,0)[bl]{...}}

\put(26,20){\makebox(0,0)[bl]{...}}


\put(35,29){\oval(10,07)}

\put(31.4,29.5){\makebox(0,0)[bl]{Disk }}
\put(34.4,26.5){\makebox(0,0)[bl]{\(\gamma\)}}

\put(30,15){\line(1,0){10}}

\put(30,15){\line(0,1){10}} \put(40,15){\line(0,1){10}}

\put(30,17){\line(1,0){10}} \put(30,19){\line(1,0){10}}
\put(30,21){\line(1,0){10}} \put(30,23){\line(1,0){10}}

\put(31,11){\makebox(0,0)[bl]{Bin \(\gamma\) }}

\put(48,78){\makebox(0,0)[bl]{\( \Longrightarrow \)}}

\put(48,64.5){\makebox(0,0)[bl]{\( \Longrightarrow \)}}

\put(48,27){\makebox(0,0)[bl]{\( \Longrightarrow \)}}

\put(80,78.5){\oval(40,09)}

\put(64,79.3){\makebox(0,0)[bl]{Computing tasks as}}
\put(62.5,74.7){\makebox(0,0)[bl]{digraph
 \(D^{2} = (T^{2},R^{2} )\)}}


\put(80,74){\vector(0,-1){4}} \put(80,70){\vector(0,1){4}}

\put(60.5,65.5){\makebox(0,0)[bl]{Data file processing graph}}
\put(65,61){\makebox(0,0)[bl]{\(G^{2}
 = <Q^{2},E_{1}^{2},E_{2}^{2} >\)}}

\put(60,60){\line(1,0){41}} \put(60,70){\line(1,0){41}}
\put(60,60){\line(0,1){10}} \put(101,60){\line(0,1){10}}

\put(80,60){\vector(0,-1){4}} \put(80,56){\vector(0,1){4}}

\put(72.5,52){\makebox(0,0)[bl]{Processor }}

\put(70,50){\line(1,0){20}} \put(70,56){\line(1,0){20}}
\put(70,50){\line(0,1){06}} \put(90,50){\line(0,1){06}}

\put(80,49.4){\vector(0,-1){4}} \put(80,45.4){\vector(0,1){4}}

\put(72,41){\makebox(0,0)[bl]{Operation}}
\put(73.8,38){\makebox(0,0)[bl]{memory }}

\put(80,41){\oval(20,08)}

\put(80,37){\vector(-3,-1){12}} \put(80,37){\vector(0,-1){4}}
\put(80,37){\vector(3,-1){12}}

\put(65,29){\oval(10,07)}

\put(61.4,29.5){\makebox(0,0)[bl]{Disk }}
\put(64.4,26.5){\makebox(0,0)[bl]{\(1\)}}

\put(60,15){\line(1,0){10}}

\put(60,15){\line(0,1){10}} \put(70,15){\line(0,1){10}}

\put(60,17){\line(1,0){10}} \put(60,19){\line(1,0){10}}
\put(60,21){\line(1,0){10}} \put(60,24){\line(1,0){10}}

\put(61,11){\makebox(0,0)[bl]{Bin \(1\) }}


\put(71,29){\makebox(0,0)[bl]{...}}

\put(71,20){\makebox(0,0)[bl]{...}}

\put(80,29){\oval(10,07)}

\put(76.4,29.5){\makebox(0,0)[bl]{Disk }}
\put(79.4,26.5){\makebox(0,0)[bl]{\(\xi\)}}

\put(75,15){\line(1,0){10}}

\put(75,15){\line(0,1){10}} \put(85,15){\line(0,1){10}}

\put(75,16){\line(1,0){10}} \put(75,19){\line(1,0){10}}
\put(75,22){\line(1,0){10}} \put(75,24){\line(1,0){10}}

\put(76,11){\makebox(0,0)[bl]{Bin \(\xi\) }}


\put(86,29){\makebox(0,0)[bl]{...}}

\put(86,20){\makebox(0,0)[bl]{...}}


\put(95,29){\oval(10,07)}

\put(91.4,29.5){\makebox(0,0)[bl]{Disk }}
\put(94.4,26.5){\makebox(0,0)[bl]{\(\gamma\)}}

\put(90,15){\line(1,0){10}}

\put(90,15){\line(0,1){10}} \put(100,15){\line(0,1){10}}

\put(90,18){\line(1,0){10}} \put(90,20){\line(1,0){10}}
\put(90,22){\line(1,0){10}} \put(90,24){\line(1,0){10}}

\put(91,11){\makebox(0,0)[bl]{Bin \(\gamma\) }}

\end{picture}
\end{center}

 Here the following notations are used:

 1. Two time moments: \(t_{1}\) and \(t_{2}\) (\(t_{2} > t_{1}\)).

 2. Digraphs over computing tasks (for \(t_{1}\) and \(t_{2}\)):

   (a) for \(t_{1}\):
  digraph \(D^{1} = (T^{1},R^{1} )\),
  where \(T^{1}\) is the set of tasks,
  \(R^{1}\) is the precedence relation as a set of arcs over the tasks
  above;

 (b) for \(t_{2}\):  \(D^{2} = (T^{2},R^{2} )\)
 (components are analogical ones);

 (c) \(T^{1}, T^{2} \subseteq T \), \(T\) is the general set of
 tasks.

 3. Data files  processing graphs
 (for \(t_{1}\) and \(t_{2}\)):

 (a) for \(t_{1}\):
 graph \(G^{1} = <Q^{1},E_{1}^{1},E_{2}^{1} >\),
 where \(Q^{1}\) is the set of files under processing at the time
 \(t_{1}\),
  \(E^{1}_{1}\) is the binary relation as precedence over the files t (i.e., a set of
 arcs),
 \(E^{1}_{2}\) is the binary relation of concurrent processing
 over the files
  (i.e., a set of edges);

 (b) for \(t_{2}\):
 \(G^{2} = <Q^{2},E_{1}^{2},E_{2}^{2} >\)
 (components are analogical ones);

(c) \(Q^{1}, Q^{2} \subseteq Q \), \(Q\) is the general set of
 files.

 4. Allocation of files (for \(t_{1}\) and \(t_{2}\)):
 \(Q\) into \(n\) disks (i.e., bins):
  \(X^{t_{1}}\),  \(X^{t_{2}}\).

 Thus, Fig. 6 illustrates re-allocation of files:~
 \(\overline{X}^{t_{1}} \Rightarrow \overline{X}^{t_{2}}\).

\subsection{Basic simple ideas for solving schemes}

 The basic simplified ideas for file allocation are the following
 (e.g., \cite{gur78,levm79}):

 1. Small and interconnected files can by integrated (condensing)
 (this leads to reduction of the problem dimension).

 2. Interconnected files have to be located on different disks
 (this leads to parallel processing without movement of hard disk heads).

 3. Interconnection relations can be integrated into a total
 integrated relations and this relation is a basis to detect
 interconnected  components in graph over files as
 cliques or quasi-cliques (communities).
 It is reasonable to obtain the communities with cardinalities
  \( \leq \gamma\) (this is the number of disks/bins).
 Thus, the examined problem consists in
  partitioning the initial graph over files into
 ``good'' interconnected subgraph.

 4. For each file community
 it is reasonable to locate its elements into different bins.

 5. Local optimization techniques can be used to improve the
 obtained solution.

 Clearly,
 the solving framework (metaheuristic)
 can be based on the ideas.





%
%




\subsection{Example of file allocation}

%
 The simplified numerical example is depicted in Fig. 7
 (as packing of items/files into disks/bins):
 (i) 8 data files~  \(Q^{1} = \{1,2,3,4,5,6,7,8\}\),
 (ii) three disks (bins),
 (iii) data processing graph~
 \(G^{1} = < Q^{1},E^{1}_{1},E^{2}_{2}>\)
 where
    relations~ \(E^{1}_{1}\), \(E^{1}_{2}\)
     are presented in Table 1 and Table 2.

\begin{center}
\begin{picture}(55,71)

\put(09,00){\makebox(0,0)[bl] {Fig. 7.
  File  location}}


\put(12,64){\circle*{1.4}}

\put(13,64){\vector(1,0){4}}

\put(20,64){\oval(06,04)} \put(19,62.5){\makebox(0,0)[bl]{\(1\)}}

\put(23,64){\vector(1,1){4}} \put(23,64){\vector(1,-1){4}}

\put(30,68){\oval(06,04)} \put(29,66.5){\makebox(0,0)[bl]{\(2\)}}

\put(30,60){\oval(06,04)} \put(29,58.5){\makebox(0,0)[bl]{\(3\)}}

\put(30,62){\line(0,1){4}}

\put(33,68){\vector(1,-1){4}}\put(33,60){\vector(1,1){4}}

\put(38,64){\circle*{1.4}}


\put(00,53){\circle*{1.8}}

\put(01.5,54){\vector(1,1){9}} \put(01.5,53){\vector(1,0){9}}
\put(01.5,52){\vector(3,-4){9}}

\put(39.5,53){\vector(1,0){9}}

\put(39.5,39){\vector(3,4){9}} \put(39.5,63){\vector(1,-1){9}}

\put(50,53){\circle*{1.8}}


\put(12,53){\circle*{1.4}}  \put(13,53){\vector(1,0){4}}

\put(23,53){\vector(1,0){4}}

\put(38,53){\circle*{1.4}}  \put(33,53){\vector(1,0){4}}

\put(20,53){\oval(06,04)} \put(19,51.5){\makebox(0,0)[bl]{\(4\)}}

\put(30,53){\oval(06,04)} \put(29,51.5){\makebox(0,0)[bl]{\(5\)}}

\put(12,38){\circle*{1.4}}

\put(13,38){\vector(1,1){8}} \put(13,38){\vector(1,0){8}}
\put(13,38){\vector(1,-1){8}}

\put(38,38){\circle*{1.4}}

\put(29,30){\vector(1,1){8}} \put(29,38){\vector(1,0){8}}
\put(29,46){\vector(1,-1){8}}


\put(30,40.5){\line(-1,2){2}}

\put(30,35.5){\line(0,1){5}}

\put(30,35.5){\line(-1,-2){2}}

\put(25,46){\oval(06,04)} \put(24,44.5){\makebox(0,0)[bl]{\(6\)}}

\put(25,38){\oval(06,04)} \put(24,36.5){\makebox(0,0)[bl]{\(7\)}}

\put(25,30){\oval(06,04)} \put(24,28.5){\makebox(0,0)[bl]{\(8\)}}

\put(25,32){\line(0,1){04}} \put(25,40){\line(0,1){04}}

\put(00,09){\line(1,0){10}}

\put(00,09){\line(0,1){17}} \put(10,09){\line(0,1){17}}

\put(00,14){\line(1,0){10}} \put(00,19){\line(1,0){10}}
\put(00,24){\line(1,0){10}}

\put(00.5,25){\line(1,0){04}} \put(05.5,25){\line(1,0){04}}

\put(0.6,20){\makebox(0,0)[bl]{File \(4\) }}
\put(0.6,15){\makebox(0,0)[bl]{File \(6\) }}
\put(0.6,10){\makebox(0,0)[bl]{File \(1\) }}

\put(01,05){\makebox(0,0)[bl]{Bin \(1\) }}

\put(20,09){\line(1,0){10}}

\put(20,09){\line(0,1){17}} \put(30,09){\line(0,1){17}}

\put(20,14){\line(1,0){10}} \put(20,19){\line(1,0){10}}
\put(20,24){\line(1,0){10}}

\put(20.5,25){\line(1,0){04}} \put(25.5,25){\line(1,0){04}}

\put(20.6,20){\makebox(0,0)[bl]{File \(5\) }}
\put(20.6,15){\makebox(0,0)[bl]{File \(7\) }}
\put(20.6,10){\makebox(0,0)[bl]{File \(2\) }}

\put(21,05){\makebox(0,0)[bl]{Bin \(2\) }}

\put(40,09){\line(1,0){10}}

\put(40,09){\line(0,1){17}} \put(50,09){\line(0,1){17}}

\put(40,14){\line(1,0){10}} \put(40,19){\line(1,0){10}}

\put(40.5,25){\line(1,0){04}} \put(45.5,25){\line(1,0){04}}

\put(40.6,15){\makebox(0,0)[bl]{File \(8\) }}
\put(40.6,10){\makebox(0,0)[bl]{File \(3\) }}

\put(41,05){\makebox(0,0)[bl]{Bin \(3\) }}

\end{picture}
\end{center}

%
  For the simplicity,
  the following is assumed:
  (a) file sizes are equals,
  (b) ordering the files at the same disk is not considered,
  (c) the cost of disk head movement from one file to another file
  at the same disk equals \(1.0\),
 (d) probabilities of movement from one file to another  file,
 initiated by processing graph, are equal.
 Table 3 contains integrated relation~ \(E^{1}_{3}\).

\begin{center}
 {\bf Table 1.} Precedence relation  \(E^{1}_{1}\) \\
\begin{tabular}{|c|  cccc  cccc|}
\hline
 \(i_{1}\)/\(i_{2}\) &\(1\)&\(2\)&\(3\)&\(4\)&\(5\)&\(6\)&\(7\)&\(8\)\\
\hline

 \(1\)&  \(\star\)&\(1\)&\(1\)&   &&&&\\
 \(2\)&\(-1\) & \(\star\)&&   &&&&\\
 \(3\)& \(-1\)&&  \(\star\)&   &&&&\\
 \(4\)& &&& \(\star\)   &\(1\)&&&\\
 \(5\)& &&&\(-1\)& \(\star\)   &&&\\
 \(6\)& &&&&& \(\star\)   &&\\
 \(7\)& &&&&&& \(\star\)   &\\
 \(8\)& &&&&&&& \(\star\)  \\

\hline
\end{tabular}
\end{center}

\begin{center}
 {\bf Table 2.} Concurrency relation \(E^{1}_{2}\) \\
\begin{tabular}{|c|  cccc  cccc|}
\hline
 \(i_{1}\)/\(i_{2}\) &\(1\)&\(2\)&\(3\)&\(4\)&\(5\)&\(6\)&\(7\)&\(8\)\\
\hline

 \(1\)&  \(\star\)&&&   &&&&\\
 \(2\)& & \(\star\)&\(1\)&   &&&&\\
 \(3\)&&\(1\)&  \(\star\)&   &&&&\\
 \(4\)& &&& \(\star\)   &&&&\\
 \(5\)& &&&& \(\star\)   &&&\\
 \(6\)& &&&&& \(\star\)   &\(1\)&\(1\)\\
 \(7\)& &&&&&\(1\)& \(\star\)   &\(1\)\\
 \(8\)& &&&&&\(1\)&\(1\)& \(\star\)  \\

\hline
\end{tabular}
\end{center}

\begin{center}
 {\bf Table 3.} Integrated relation  \(E^{1}_{3}\) \\
\begin{tabular}{|c|  cccc  cccc|}
\hline
 \(i_{1}\)/\(i_{2}\) &\(1\)&\(2\)&\(3\)&\(4\)&\(5\)&\(6\)&\(7\)&\(8\)\\
\hline

 \(1\)&  \(\star\)&\(1\)&\(1\)&   &&&&\\
 \(2\)&\(1\) & \(\star\)&\(1\)&   &&&&\\
 \(3\)&\(1\)&\(1\)&  \(\star\)&   &&&&\\
 \(4\)& &&& \(\star\)   &\(1\)&&&\\
 \(5\)& &&&\(1\)& \(\star\)   &&&\\
 \(6\)& &&&&& \(\star\)   &\(1\)&\(1\)\\
 \(7\)& &&&&&\(1\)& \(\star\)   &\(1\)\\
 \(8\)& &&&&&\(1\)&\(1\)& \(\star\)  \\

\hline
\end{tabular}
\end{center}

 Relation
 \(E^{1}_{3}\)
 is a basis to detect \(3\)
 interconnected components as cliques or quasi-cliques
 (by processing) as follows:
 \(\{1,2,3 \}\),
 \(\{6,7,8 \}\), \(\{4,5 \}\).

 Further, it is reasonable
 to locate elements of each clique above into different bins/disks.
 Thus, the file allocation solution  is
 (without file ordering on each disk) (Fig. 7):
 \(X^{1}_{1} = \{1,4,6\}\),
 \(X^{1}_{2} = \{2,5,7\}\),
  \(X^{1}_{3} = \{3,8\}\),
  i.e.,
 \(X^{1} = \{X^{1}_{1},X^{1}_{2},X^{1}_{3} \}\)
  and
  the corresponding value of objective function is:~
 \(\Psi ( \overline{X^{1}} )  = 0\).

\section{Reconfiguration (restructuring) of file allocation solutions}

 In this section, two problems of
 solution reconfiguration are described:
  one-stage restructuring
   \cite{lev11restr,lev15restr,shieb12}
 and two-stage restructuring
 \cite{lev11restr,lev15restr}.
 This restructuring approach
 is applied for
 data file allocation solutions
 (i.e., reconfiguration of allocation solutions).
 It is assumed
 the cost of file relocation  operation  from one disk to another disk
 is equal \(1.0\).
%
%
 The allocation problem from previous section is considered as the
 stage 1 (\(t=t_{1}\))
  with corresponding allocation solution \(X^{1}\) (Fig. 7).

\subsection{One-stage restructuring}

 Here a next time stage (stage 2, \(t=t_{2}\)) is considered (Fig. 8, Table 4, Table 5).
 Integrated relation over files is contained in Table 6.
  \(3\) interconnected components are:
 \(\{1,4,5 \}\),
 \(\{2,3,6 \}\), \(\{7,8 \}\).

 The corresponding solution is   (Fig. 8, \(t=t_{2}\)):
 \(X^{2}_{1} = \{1,2,7\}\),
 \(X^{2}_{2} = \{4,3,8\}\),
  \(X^{2}_{3} = \{5,6\}\);
 i.e,
 \(X^{2} = \{X^{2}_{1},X^{2}_{2},X^{2}_{3} \}\)
  and
  the corresponding value of objective function is:~
 \(\Psi ( \overline{X^{2}} )  = 0\).

 Thus, the following restructuring problem is examined
 (ordering of file on disk is not considered)
 \cite{lev11restr,lev15restr}:

~~

  Modify solution \(X^{1}\) into restructured solution \(X^{2*}\)
 such that
    \[ \min~ \rho (X^{2*},X^{2}) = |\Phi(X^{2*})-\Phi (X^{2})|
   ~(proximity)
      ~~~~~~ s.t.~~~  h (X^{1}  \Rightarrow X^{2*})  \leq 2.0 ~(modification~cost).\]
%

\begin{center}
\begin{picture}(55,58)

\put(03,00){\makebox(0,0)[bl] {Fig. 8.
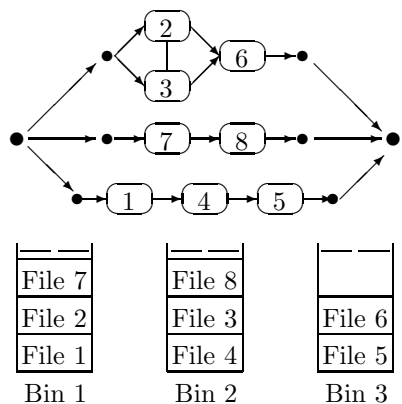
  File  location (\(t=t_{2}\))}}


\put(12,51){\circle*{1.4}}

\put(13,51){\vector(1,1){4}} \put(13,51){\vector(1,-1){4}}

\put(30,51){\oval(06,04)} \put(29,49.5){\makebox(0,0)[bl]{\(6\)}}

\put(33,51){\vector(1,0){4}}

\put(20,55){\oval(06,04)} \put(19,53.5){\makebox(0,0)[bl]{\(2\)}}

\put(20,47){\oval(06,04)} \put(19,45.5){\makebox(0,0)[bl]{\(3\)}}

\put(20,49){\line(0,1){4}}

\put(23,55){\vector(1,-1){4}}\put(23,47){\vector(1,1){4}}

\put(38,51){\circle*{1.4}}


\put(00,40){\circle*{1.8}}

\put(01.5,41){\vector(1,1){9}} \put(01.5,40){\vector(1,0){9}}

\put(01.5,39){\vector(1,-1){6}}

\put(39.5,40){\vector(1,0){9}}

\put(43,32){\vector(1,1){5.8}}

\put(39.5,50){\vector(1,-1){9}}

\put(50,40){\circle*{1.8}}


\put(12,40){\circle*{1.4}}  \put(13,40){\vector(1,0){4}}

\put(23,40){\vector(1,0){4}}

\put(38,40){\circle*{1.4}}  \put(33,40){\vector(1,0){4}}

\put(20,40){\oval(06,04)} \put(19,38.5){\makebox(0,0)[bl]{\(7\)}}

\put(30,40){\oval(06,04)} \put(29,38.5){\makebox(0,0)[bl]{\(8\)}}

\put(08,32){\circle*{1.4}}

\put(08,32){\vector(1,0){4}} \put(38,32){\vector(1,0){4}}

\put(42,32){\circle*{1.4}}

\put(15,32){\oval(06,04)} \put(14,30.5){\makebox(0,0)[bl]{\(1\)}}

\put(25,32){\oval(06,04)} \put(24,30.5){\makebox(0,0)[bl]{\(4\)}}

\put(35,32){\oval(06,04)} \put(34,30.5){\makebox(0,0)[bl]{\(5\)}}

\put(18,32){\vector(1,0){04}} \put(28,32){\vector(1,0){04}}

\put(00,09){\line(1,0){10}}

\put(00,09){\line(0,1){17}} \put(10,09){\line(0,1){17}}

\put(00,14){\line(1,0){10}} \put(00,19){\line(1,0){10}}
\put(00,24){\line(1,0){10}}

\put(00.5,25){\line(1,0){04}} \put(05.5,25){\line(1,0){04}}

\put(0.6,20){\makebox(0,0)[bl]{File \(7\) }}
\put(0.6,15){\makebox(0,0)[bl]{File \(2\) }}
\put(0.6,10){\makebox(0,0)[bl]{File \(1\) }}

\put(01,05){\makebox(0,0)[bl]{Bin \(1\) }}

\put(20,09){\line(1,0){10}}

\put(20,09){\line(0,1){17}} \put(30,09){\line(0,1){17}}

\put(20,14){\line(1,0){10}} \put(20,19){\line(1,0){10}}
\put(20,24){\line(1,0){10}}

\put(20.5,25){\line(1,0){04}} \put(25.5,25){\line(1,0){04}}

\put(20.6,20){\makebox(0,0)[bl]{File \(8\) }}
\put(20.6,15){\makebox(0,0)[bl]{File \(3\) }}
\put(20.6,10){\makebox(0,0)[bl]{File \(4\) }}

\put(21,05){\makebox(0,0)[bl]{Bin \(2\) }}

\put(40,09){\line(1,0){10}}

\put(40,09){\line(0,1){17}} \put(50,09){\line(0,1){17}}

\put(40,14){\line(1,0){10}} \put(40,19){\line(1,0){10}}

\put(40.5,25){\line(1,0){04}} \put(45.5,25){\line(1,0){04}}

\put(40.6,15){\makebox(0,0)[bl]{File \(6\) }}
\put(40.6,10){\makebox(0,0)[bl]{File \(5\) }}

\put(41,05){\makebox(0,0)[bl]{Bin \(3\) }}

\end{picture}
\end{center}

\begin{center}
 {\bf Table 4.} Precedence relation  \(E^{2}_{1}\) \\
\begin{tabular}{|c|  cccc  cccc|}
\hline
 \(i_{1}\)/\(i_{2}\) &\(1\)&\(2\)&\(3\)&\(4\)&\(5\)&\(6\)&\(7\)&\(8\)\\
\hline

 \(1\)& \(\star\)&&& \(1\)  &\(1\)&&&\\
 \(2\)& & \(\star\)&&   &&\(1\)&&\\
 \(3\)& &&  \(\star\)&  &&\(1\)&&\\
 \(4\)&\(-1\) &&& \(\star\)   &\(1\)&&&\\
 \(5\)& \(-1\)&&&\(-1\)& \(\star\)   &&&\\
 \(6\)& &\(-1\)&\(-1\)&&& \(\star\)   &&\\
 \(7\)& &&&&&& \(\star\)   &\(1\)\\
 \(8\)& &&&&&&\(-1\)& \(\star\)  \\

\hline
\end{tabular}
\end{center}

\begin{center}
 {\bf Table 5.} Concurrency relation  \(E^{2}_{2}\) \\
\begin{tabular}{|c|  cccc  cccc|}
\hline
 \(i_{1}\)/\(i_{2}\) &\(1\)&\(2\)&\(3\)&\(4\)&\(5\)&\(6\)&\(7\)&\(8\)\\
\hline

 \(1\)& \(\star\) &&&   &&&&\\
 \(2\)& & \(\star\)&\(1\)&   &&&&\\
 \(3\)&&\(1\)&  \(\star\)&   &&&&\\
 \(4\)& &&& \(\star\)   &&&&\\
 \(5\)& &&&& \(\star\)   &&&\\
 \(6\)& &&&&& \(\star\)   &&\\
 \(7\)& &&&&&& \(\star\)  &\\
 \(8\)& &&&&&&& \(\star\)  \\

\hline
\end{tabular}
\end{center}

\begin{center}
 {\bf Table 6.} Integrated relation  \(E^{2}_{3}\) \\
\begin{tabular}{|c|  cccc  cccc|}
\hline
 \(i_{1}\)/\(i_{2}\) &\(1\)&\(2\)&\(3\)&\(4\)&\(5\)&\(6\)&\(7\)&\(8\)\\
\hline

 \(1\)& \(\star\)&&& \(1\)  &\(1\)&&&\\
 \(2\)& & \(\star\)&\(1\)&   &&\(1\)&&\\
 \(3\)& &\(1\)&  \(\star\)&  &&\(1\)&&\\
 \(4\)&\(1\) &&& \(\star\)   &\(1\)&&&\\
 \(5\)& \(1\)&&&\(1\)& \(\star\)   &&&\\
 \(6\)& &\(1\)&\(1\)&&& \(\star\)   &&\\
 \(7\)& &&&&&& \(\star\)   &\(1\)\\
 \(8\)& &&&&&&\(1\)& \(\star\)  \\

\hline
\end{tabular}
\end{center}

 Note, restructuring  process  \(X^{1} \Rightarrow X^{2}\)
 consists of the following file re-locations operations:
 (i) file \(4\) is relocated from disk \(1\) into disk \(3\),
 (ii) file \(5\) is relocated from disk \(2\) into disk \(1\),
 and
 (iii) file \(1\) is relocated from disk \(1\) into disk \(2\).
 The cost of the relocation problem is:~
    \( h (X^{1}  \Rightarrow X^{2}) = 3.0\).

 Now the following restructuring process for \(X^{1} \Rightarrow X^{2*}\)
 is examined:
 (i) file \(5\) is relocated from disk \(2\) into disk \(1\),
 (ii) file \(1\) is relocated from disk \(1\) into disk \(2\).
 The obtained restructured solution is:~

 \(X^{2*} = \{X^{2*}_{1},  X^{2*}_{2}, X^{2*}_{3} \}\) where
  \(X^{2*}_{1} = \{4,5,6   \}\),
 \(X^{2*}_{1} = \{1,2,3   \}\),
 \(X^{2*}_{3} = \{3,8 \}\).

 The cost of the relocation problem is:~
    \( h (X^{1}  \Rightarrow X^{2*}) = 2.0\),
  the corresponding value of objective function is:~
 \(\Psi ( \overline{X^{2*}} )  = 1.0\)
 (here the disk head movement is needed
  from file \(4\) to file \(5\), \(t=t_{2}\)).

\subsection{Multistage restructuring}

 Here a next time stage (stage 3, \(t=t_{3}\))
 is considered (Fig. 9, Table 7, Table 8).
 Integrated relation over files is contained in Table 9.
 \(3\) interconnected components are:
 \(\{1,2,4 \}\),
 \(\{3,5,6 \}\), \(\{7,8 \}\).

 The corresponding solution is (Fig. 9, \(t=t_{3}\)):
 \(X^{3}_{1} = \{1,3,7\}\),
 \(X^{3}_{2} = \{2,5,8\}\),
  \(X^{3}_{3} = \{4,6\}\),
 i.e,
 \(X^{3} = \{X^{3}_{1},X^{3}_{2},X^{3}_{3} \}\)
  and
  the corresponding value of objective function is:~
 \(\Psi ( \overline{X^{3}} )  = 0\).

\begin{center}
\begin{picture}(55,71)

\put(03,00){\makebox(0,0)[bl] {Fig. 9.
  File location (\(t=t_{3}\))}}


\put(12,64){\circle*{1.4}}

\put(13,64){\vector(1,0){4}}

\put(20,64){\oval(06,04)} \put(19,62.5){\makebox(0,0)[bl]{\(1\)}}

\put(23,64){\vector(1,1){4}} \put(23,64){\vector(1,-1){4}}

\put(30,68){\oval(06,04)} \put(29,66.5){\makebox(0,0)[bl]{\(2\)}}

\put(30,60){\oval(06,04)} \put(29,58.5){\makebox(0,0)[bl]{\(4\)}}

\put(30,62){\line(0,1){4}}

\put(33,68){\vector(1,-1){4}}\put(33,60){\vector(1,1){4}}

\put(38,64){\circle*{1.4}}


\put(00,53){\circle*{1.8}}

\put(01.5,54){\vector(1,1){9}} \put(01.5,53){\vector(1,0){9}}
\put(01.5,52){\vector(3,-4){9}}

\put(39.5,53){\vector(1,0){9}}

\put(39.5,39){\vector(3,4){9}} \put(39.5,63){\vector(1,-1){9}}

\put(50,53){\circle*{1.8}}


\put(12,53){\circle*{1.4}}  \put(13,53){\vector(1,0){4}}

\put(23,53){\vector(1,0){4}}

\put(38,53){\circle*{1.4}}  \put(33,53){\vector(1,0){4}}

\put(20,53){\oval(06,04)} \put(19,51.5){\makebox(0,0)[bl]{\(7\)}}

\put(30,53){\oval(06,04)} \put(29,51.5){\makebox(0,0)[bl]{\(8\)}}

\put(12,38){\circle*{1.4}}

\put(13,38){\vector(1,1){8}} \put(13,38){\vector(1,0){8}}
\put(13,38){\vector(1,-1){8}}

\put(38,38){\circle*{1.4}}

\put(29,30){\vector(1,1){8}} \put(29,38){\vector(1,0){8}}
\put(29,46){\vector(1,-1){8}}


\put(30,40.5){\line(-1,2){2}}

\put(30,35.5){\line(0,1){5}}

\put(30,35.5){\line(-1,-2){2}}

\put(25,46){\oval(06,04)} \put(24,44.5){\makebox(0,0)[bl]{\(3\)}}

\put(25,38){\oval(06,04)} \put(24,36.5){\makebox(0,0)[bl]{\(5\)}}

\put(25,30){\oval(06,04)} \put(24,28.5){\makebox(0,0)[bl]{\(6\)}}

\put(25,32){\line(0,1){04}} \put(25,40){\line(0,1){04}}

\put(00,09){\line(1,0){10}}

\put(00,09){\line(0,1){17}} \put(10,09){\line(0,1){17}}

\put(00,14){\line(1,0){10}} \put(00,19){\line(1,0){10}}
\put(00,24){\line(1,0){10}}

\put(00.5,25){\line(1,0){04}} \put(05.5,25){\line(1,0){04}}

\put(0.6,20){\makebox(0,0)[bl]{File \(7\) }}
\put(0.6,15){\makebox(0,0)[bl]{File \(3\) }}
\put(0.6,10){\makebox(0,0)[bl]{File \(1\) }}

\put(01,05){\makebox(0,0)[bl]{Bin \(1\) }}

\put(20,09){\line(1,0){10}}

\put(20,09){\line(0,1){17}} \put(30,09){\line(0,1){17}}

\put(20,14){\line(1,0){10}} \put(20,19){\line(1,0){10}}
\put(20,24){\line(1,0){10}}

\put(20.5,25){\line(1,0){04}} \put(25.5,25){\line(1,0){04}}

\put(20.6,20){\makebox(0,0)[bl]{File \(8\) }}
\put(20.6,15){\makebox(0,0)[bl]{File \(5\) }}
\put(20.6,10){\makebox(0,0)[bl]{File \(2\) }}

\put(21,05){\makebox(0,0)[bl]{Bin \(2\) }}

\put(40,09){\line(1,0){10}}

\put(40,09){\line(0,1){17}} \put(50,09){\line(0,1){17}}

\put(40,14){\line(1,0){10}} \put(40,19){\line(1,0){10}}

\put(40.5,25){\line(1,0){04}} \put(45.5,25){\line(1,0){04}}

\put(40.6,15){\makebox(0,0)[bl]{File \(6\) }}
\put(40.6,10){\makebox(0,0)[bl]{File \(4\) }}

\put(41,05){\makebox(0,0)[bl]{Bin \(3\) }}

\end{picture}
\end{center}

\begin{center}
 {\bf Table 7.} Precedence relation  \(E^{3}_{1}\) \\
\begin{tabular}{|c|  cccc  cccc|}
\hline
 \(i_{1}\)/\(i_{2}\) &\(1\)&\(2\)&\(3\)&\(4\)&\(5\)&\(6\)&\(7\)&\(8\)\\
\hline

 \(1\)& \(\star\)&\(1\)&&\(1\)   &&&&\\
 \(2\)&\(-1\) & \(\star\)&&   &&&&\\
 \(3\)& &&  \(\star\)&   &&&&\\
 \(4\)&\(-1\) &&& \(\star\)   &&&&\\
 \(5\)& &&&& \(\star\)   &&&\\
 \(6\)& &&&&&  \(\star\) &&\\
 \(7\)& &&&&&& \(\star\)   & \(1\)\\
 \(8\)& &&&&&&\(-1\)& \(\star\)  \\

\hline
\end{tabular}
\end{center}

\begin{center}
 {\bf Table 8.} Concurrency relation  \(E^{3}_{2}\) \\
\begin{tabular}{|c|  cccc  cccc|}
\hline
 \(i_{1}\)/\(i_{2}\) &\(1\)&\(2\)&\(3\)&\(4\)&\(5\)&\(6\)&\(7\)&\(8\)\\
\hline

 \(1\)&  \(\star\)&&&   &&&&\\
 \(2\)& & \(\star\)&&\(1\)   &&&&\\
 \(3\)& && \(\star\)&   &\(1\)&\(1\)&&\\
 \(4\)& &\(1\)&& \(\star\)   &&&&\\
 \(5\)& &&\(1\)&& \(\star\)   &\(1\)&&\\
 \(6\)& &&\(1\)&&\(1\)& \(\star\)   &&\\
 \(7\)& &&&&&& \(\star\) &\\
 \(8\)& &&&&&&& \(\star\)  \\

\hline
\end{tabular}
\end{center}

\newpage
\begin{center}
 {\bf Table 9.} Integrated relation  \(E^{3}_{3}\) \\
\begin{tabular}{|c|  cccc  cccc|}
\hline
 \(i_{1}\)/\(i_{2}\) &\(1\)&\(2\)&\(3\)&\(4\)&\(5\)&\(6\)&\(7\)&\(8\)\\
\hline

 \(1\)&  \(\star\)&\(1\)&& \(1\)  &&&&\\
 \(2\)&\(1\) & \(\star\)&&\(1\)   &&&&\\
 \(3\)&&&  \(\star\)&   &\(1\)&\(1\)&&\\
 \(4\)&\(1\) &\(1\)&& \(\star\)   &&&&\\
 \(5\)& &&\(1\)&& \(\star\)   &\(1\)&&\\
 \(6\)& &&\(1\)&&\(1\)& \(\star\)   &&\\
 \(7\)& &&&&&& \(\star\) &\(1\)\\
 \(8\)& &&&&&&\(1\)& \(\star\)  \\

\hline
\end{tabular}
\end{center}

 Note, restructuring  process  \(X^{2} \Rightarrow X^{3}\)
 consists of the following file re-locations operations:
 (i) file \(4\) is relocated from disk \(2\) into disk \(3\),
 (ii) file \(5\) is relocated from disk \(3\) into disk \(2\),
 (iii) file \(3\) is relocated from disk \(2\) into disk \(1\),
 and
 (iv) file \(1\) is relocated from disk \(1\) into disk \(2\).
 The cost of the relocation problem is:~
    \( h (X^{2}  \Rightarrow X^{3}) = 4.0\).

 Now the following restructuring process for \(X^{2*} \Rightarrow X^{3*}\)
 is examined:
%
%
 (i) file \(3\) is relocated from disk \(3\) into disk \(1\),
 (ii) file \(2\) is relocated from disk \(2\) into disk \(3\).
 The obtained restructured solution is:~

 \(X^{3*} = \{X^{3*}_{1},  X^{3*}_{2}, X^{3*}_{3} \}\) where
  \(X^{3*}_{1} = \{4,5,6   \}\),
 \(X^{3*}_{1} = \{1,3,7   \}\),
 \(X^{3*}_{3} = \{2,8 \}\).

 The cost of the relocation problem is:~
 \( h (X^{2*}  \Rightarrow X^{3*}) = 2.0\),
  the corresponding value of objective function is:~
 \(\Psi ( \overline{X^{3*}} ) = 1.0\)
 (here the disk head movement is needed
 from file \(5\) to file \(6\), \(t=t_{3}\)).

 Finally, two 3-stage file allocation trajectory can be considered:

 (i) trajectory consisting of local optimal solutions~
 \( S^{opt} = < X^{1}, X^{2}, X^{3} > \),
  here total solution modification cost equals \(7.0\);

 (ii) trajectory consisting of restructured solutions~
 \( S^{restr} = < X^{1}, X^{2*}, X^{3*} > \),
  here total solution modification cost equals \(4.0\) and
   proximity to optimal value of objective function at stage 2 and stage 3
  will be equal \(1.0\)
  (this case corresponds to sequential solving strategy \cite{lev15restr}).

 Evidently,
 it is possible to manage the parameters of the restructuring
 process,
 i.e., by changes of the required constraint(s) for modification
 cost(s) for restructuring problems.


\section{Conclusion}

 The paper contains description of
 data allocation in two-layer computer storage
 (several disks).
 Models
 and simplified heuristics were described.
 In addition,
 solution reconfiguration problems for data allocation on disks was  suggested:
 (i) one-stage restructuring,
 (ii) multistage restructuring.
 It is necessary to point out
 other applications
 as allocation of objects into parallel resources,
 for example:
 (1) distributed computer systems (e.g., task allocation
 while taking into account tasks interconnection),
 (2) communication systems:
 (2.1)
 planning of multiple access communication channels
 (e.g., allocation of messages into subchannels
 while taking into account message interference),
  (2.2) planning of multiple beam antenna
 (e.g., allocation of messages into antenna subbeams
 while taking into account message interference),
 (2.3) connection of end-users and access points
 in communication systems.

 The prospective future research directions are the following:
 (a) examination of the suggested problems
 with different file sizes,
 (b) taking into account uncertainty in models,
%
 (c) execution of computer experiments
 for analysis and comparison of various solving methods,
 (d) consideration of other application domains,
  and
 (e) usage of the described approaches  in CS/engineering education.

\section{Acknowledgments}

 The research materials presented in the article
 were partially supported by The Russian Foundation for
 Basic Research, project 15-07-01241
 ``Reconfiguration of Solutions in Combinatorial Optimization''
 (principal investigator: Mark Sh. Levin).


\end{document}